\documentclass[11pt]{article}
\newcommand{\be}{\begin{equation}}
\newcommand{\ee}{\end{equation}}
\newcommand{\ba}{\begin{array}}
\newcommand{\ea}{\end{array}}

\newcommand{\bea}{\begin{eqnarray*}}
\newcommand{\eea}{\end{eqnarray*}}
\newcommand{\bean}{\begin{eqnarray}}
\newcommand{\eean}{\end{eqnarray}}
\newcommand{\proof}{\vspace{1ex}\noindent{\em Proof}. \ }
\def\ds{\displaystyle}
\def\nm{\noalign{\medskip}}

\newcommand{\R}{\mbox{\bf R}}

\newcommand\eps{\varepsilon}

\newtheorem{lem}{Lemma}[section]

\newtheorem{prop}{Proposition}[section]
\def\Box{\leavevmode\vbox{\hrule
     \hbox{\vrule\kern5pt\vbox{\kern5pt}%
           \vrule}\hrule}}
\newcommand{\square}{\hfill$\Box$}

\begin{document}
\title{Recovery of small electromagnetic inhomogeneities from boundary
measurements on part of the boundary}

\author{H. Ammari \thanks{ Centre de Math\'ematiques Appliqu\'ees,
CNRS UMR 7641 \& Ecole Polytechnique, 91128 Palaiseau Cedex,
France (Email: ammari@cmapx.polytechnique.fr).}   \and A. G. Ramm
\thanks{Mathematics Department, Kansas State University,
Manhattan, KS 66506-2602, USA (Email: ramm@math.ksu.edu).}}
\maketitle \abstract{ We consider for the Helmholtz equation the
inverse problem of identifying locations
 and certain properties of the shapes of small dielectric inhomogeneities
in a homogeneous background medium from boundary measurements on
part of the boundary. Using as weights particular background
solutions constructed by solving a minimization problem we develop
an asymptotic (variational) method based on appropriate averaging
of the partial boundary measurements. }

\begin{center}
{\bf Identification de petites inhomog\'en\'eit\'es
di\'electriques \`a partir de mesures sur une partie du bord}
\end{center}

\begin{center}
{\bf R\'esum\'e}
\end{center}
Nous consid\'erons le probl\`eme d'identification de petites
inhomog\'en\'eit\'es di\'electriques \`a partir de mesures
(incompl\`etes) sur uniquement une partie du bord. Gr\^ace \`a un
r\'esultat de densit\'e, nous construisons une  fonction  dont
la transform\'ee de Fourier inverse permet de localiser les
petites inhomog\'en\'eit\'es.

\section*{Version fran\c caise abr\'eg\'ee} Soit $\Omega$ un ouvert
born\'e  de $\R^d$, $d\geq 2$, de classe $C^2$. Supposons que  $\Omega$
contient $m$ inhomog\'en\'eit\'es,$\{z_j + \alpha B_j\}_{j=1}^m$,
o\`u $\alpha$ est un petit param\`etre, $B_j \subset \R^d$ est un
ouvert born\'e et les points $\{z_j\}_{j=1}^m$ v\'erifient les
hypoth\`eses (\ref{f1}). Soient la perm\'eabilit\'e magn\'etique
$\mu_\alpha$ et la permittivit\'e \'electrique $\eps_\alpha$ de
forme (\ref{murhodef}).

Cette Note concerne le probl\`eme de reconstruction des points
$\{z_j\}_{j=1}^m$ et des tenseurs de polarization $M_j$ des
domaines $B_j$, d\'efinis par (\ref{Meq}), \`a partir de la mesure
de la d\'eriv\'ee normale du champ \'electrique $E_\alpha$,
solution de l'\'equation de Helmholtz (\ref{f2}), sur une partie
$\Gamma_1 \subset \subset \partial \Omega$. Gr\^ace \`a un
r\'esultat de densit\'e, \'etabli dans la Proposition \ref{lem2}, 
et \`a la formule asymptotique (\ref{eq1}), nous r\'eduisons ce
probl\`eme inverse au calcul de la transform\'ee de Fourier
inverse de la fonction $\Lambda_\alpha$, d\'efinie dans
(\ref{f3}). La fonction test $w_\alpha$,  utilis\'ee dans (\ref{f3}),
peut \^etre construite num\'eriquement en r\'esolvant le probl\`eme 
de minimisation
(\ref{f4}). Cette fonction v\'erifie les propri\'et\'es
d'approximation \'ennonc\'ees dans le Lemme \ref{lem1}. Cette Note
g\'en\'eralise la m\'ethode introduite dans \cite{AMV} aux
situations o\`u on ne disposerait pas de la mesure de la
d\'eriv\'ee normale du champ \'electrique $E_\alpha$ sur tout le
bord $\partial \Omega$.

\section{Introduction} Let $\Omega$ be a bounded $C^2$-domain
in $\R^d$, $d\geq 2$ and
 $\nu$ be the outward unit normal to
$\partial \Omega$. Assume  that $\Omega$ contains a finite
number of inhomogeneities, each of the form $z_j + \alpha B_j$,
where $B_j \subset \R^d$ is a bounded, smooth domain containing
the origin. The total collection of inhomogeneities is
 $ \ds {\cal B}_\alpha
 = \ds \cup_{j=1}^{m} (z_j  + \alpha B_j)$.
  The points $z_j \in \Omega, j=1, \ldots, m,$ which determine the
  location of the inhomogeneities, are assumed to satisfy the
  following inequalities:
\be
\label{f1}
 | z_j  - z_l | \geq c_0 > 0, \forall \; j \neq l  \quad
\mbox{ and } \mbox{ dist} (z_j, \partial \Omega) \geq c_0
> 0, \forall \; j.
\ee This assumption implies
 $m \leq \ds \frac{2 d |\Omega|}{\pi c_0^d}$. Assume
that $\alpha
>0$, the common order of magnitude of the diameters of the
inhomogeneities, is sufficiently small, that these
 inhomogeneities are disjoint,  and that
their distance to $\R^d \setminus \overline{\Omega}$ is larger than
$c_0/2$.  Let $\mu_0$ and $\eps_0$ denote the permeability and
the permittivity of the background medium, and assume that
{\it $\mu_0>0$ and $\eps_0>0$ are positive constants}.  Let $\mu_j>0$ and
$\eps_j>0$
denote the permeability and the permittivity of the j-th
inhomogeneity, $z_j+\alpha B_j$, these are also assumed to be
positive constants. Introduce the piecewise-constant magnetic permeability
\begin{equation}
\mu_\alpha(x)=\left \{ \begin{array}{*{2}{l}}
 \mu_0,\;\;& x \in \Omega \setminus \bar {\cal B}_\alpha,  \\
 \mu_j,\;\;& x \in z_j+\alpha B_j, \;j=1 \ldots m.
\end{array}
\right . \label{murhodef}
\end{equation}
If we allow the degenerate case $\alpha =0$, then the function
$\mu_0(x)$ equals the constant $\mu_0$. The piecewise constant
electric permittivity, $\eps_\alpha(x)$ is defined analogously.

Consider solutions to the time-harmonic Maxwell's equations
with $\exp (-i \omega t)$ time dependence.
 Let $E_\alpha$ be the electric field
in the presence of the inhomogeneities. It solves the Helmholtz
equation $$ \nabla \cdot (\frac{1}{\mu_\alpha} \nabla E_\alpha) +
\omega^2 \eps_\alpha E_\alpha = 0 \ \ \mbox{in } \Omega,$$ with
the boundary condition $E_\alpha = f$ on $\partial \Omega$,  where
$\omega > 0$ is a given frequency. The electric field, $E_0$, in
the absence of any inhomogeneities, satisfies the
following equation:
\be
\label{f2}
 \Delta E_0 + k^2 E_0 = 0  \ \ \mbox{in } \Omega,
 \ee
where $k^2 = \omega^2 \mu_0 \eps_0$, with $E_0 = f$ on $\partial
\Omega$. In order to insure well-posedness (also for the
$\alpha$-dependent case for $\alpha$ sufficiently small) we shall
assume that $k^2$ is not an eigenvalue for the operator $- \Delta$
in $L^2(\Omega)$
with the Dirichlet boundary conditions. It has been shown
in \cite{VV} that the following asymptotic formula  holds
uniformly on $\partial \Omega$ \be \label{eq1}
\begin{array}{l}
\ds \frac{\partial E_\alpha}{\partial \nu}(x) - \frac{\partial
E_0}{\partial \nu}(x) - 2 \int_{\partial \Omega}(\frac{\partial
E_\alpha}{\partial \nu}- \frac{\partial E_0}{\partial \nu})(y)
\frac{\partial G(x, y)}{\partial \nu} \ ds(y) \\ \nm \ \ \ \ \ds =
2 \alpha^d \sum_{j=1}^m (1 - \frac{\mu_0}{\mu_j}) \nabla_y
\frac{\partial G(x, z_j)}{\partial \nu(x)} \cdot
M_j(\frac{\mu_j}{\mu_0}) \nabla E_0(z_j) \\
 \nm \ \ \ \ \ds - 2 \alpha^d k^2 \sum_{j=1}^m (1 - \frac{\eps_j}{\eps_0})
\frac{\partial G(x, z_j)}{\partial \nu(x)} |B_j| E_0(z_j) +
o(\alpha^d),
\end{array}
\ee where the remainder $o(\alpha^d)$ is independent of the set of
points $\{z_j\}_{j=1}^m$ provided that (\ref{f1}) holds, $G(x, y)$
is a free space Green's function for $\Delta + k^2$, and each
$M_j$ is a $d \times d$, symmetric, positive definite matrix
associated with the j-th inhomogeneity, called the polarizability
tensor, which is given by
\begin{equation}
(M_j)_{ll^\prime}=|B_j|\delta_{ll^\prime}+
\left(\frac{\mu_j}{\mu_0} -1\right) \int_{\partial
B_j}y_l{\partial\phi_{l^\prime}^+\over{
\partial\nu_j}}d\sigma_y,\label{Meq}
\end{equation}
where, for $1 \leq l^\prime \leq d$, $\phi_{l^\prime}(y)$ is the
unique function which satisfies \[ \left\{ \begin{array}{lll}
 \Delta\phi_{l^\prime} &=& 0\ \ \ \ \mbox{in}\ \ \ B_j \ \ \mbox{and} \ \ \ \R^d \setminus
\overline{B_j},\\ \nm \ds \frac{1}{\mu_0}  {\partial
\phi_{l^\prime}^-\over{\partial\nu_j}}- \frac{1}{\mu_j}
{\partial\phi_{l^\prime}^+\over{\partial\nu_j}} &=& \ds -
\frac{1}{\mu_j} \nu_j \cdot e_{l^\prime} \ \ \ \ \mbox{on}\ \ \
\partial B_j,
\end{array}
\right.
\]
with $\phi_{l^\prime}$ continuous across $\partial B_j$ and
$\lim_{|y|\rightarrow\infty} \phi_{l^\prime}(y)=0.$ Here
$\{e_{l^\prime}\}_{l^\prime =1}^d$ is an orthonormal basis of
$\R^d$, $\nu_j$ denotes the outward unit normal to $\partial B_j$,
superscripts $-$ and $+$ indicate the limiting values as the point
approaches
$
\partial B_j$ from outside $B_j$, and from
inside $B_j$, respectively.

Our goal is to identify the locations
$\{z_j\}_{j=1}^m$ and the polarizability tensors $\{M_j\}_{j=1}^m$
of the small inhomogeneities ${\cal B}_\alpha$ from the boundary
measurements of $\frac{\partial E_\alpha}{\partial \nu}$ on a
given part $\Gamma_1 \subset \subset \partial \Omega$. This
inverse problem is more complicated from the mathematical point of
view and more interesting in applications
than the one solved in \cite{AMV}, because
in many applications
one cannot get measurements on the whole boundary.
As in \cite{AMV}, we
want to reduce this reconstruction problem to the calculation of
an inverse Fourier transform. Our work is
a generalization of \cite{AMV} to the case of the reconstruction
of
small inhomogeneities from the measurements on a part of the boundary.

In  \cite{395} a different method is proposed for finding small
inhomogeneities from the scattering data. In  \cite{396} it is proved
that the singular (delta-type) potentials are uniquely determined
by the scattering data. In \cite{399} a numerical realization of the
method proposed in \cite{395} is given. In \cite{364} a method for
finding small inhomogeneities from tomographic data is given. In
\cite{144} analytic formulas are derived for electric and magnetic
polarizability tensors for bodies of arbitrary shapes. In \cite{278}
a number of multi-dimensional inverse scattering problems are
studied and some of the methods developed there are related to our
approach.

\section{Identification procedure}
Before describing our identification procedure,  let us introduce
the sets $N(\Omega) = \{v: v \in H^1(\overline{\Omega}) \cap H^2(\Omega),
\Delta v + k^2 v = 0 \
\mbox{in} \ \Omega  \}$ and $\tilde{N}(\Omega) = \{v: v \in
 H^1(\overline{\Omega}) \cap H^2(\Omega),
\Delta v + k^2 v = 0 \ \mbox{in} \ \Omega, v= 0 \
\mbox{on} \ \ \Gamma_2 \}$, where $\Gamma_2
=
\partial \Omega \setminus \overline{\Gamma_1}$, where
$\Gamma_1$ is an open in $\partial \Omega$ subset.

The general
approach we use to recover the locations and the
polarizability  tensors of the small inhomogeneities is to integrate
the solution $E_\alpha$ against special test functions in the set
$\tilde{N}(\Omega)$. Let $v$ be any function  in
$\tilde{N}(\Omega)$.
  As in \cite{AMV}, the following estimate can
be derived from (\ref{eq1}): \be \label{eq3}
\begin{array}{lll} \ds
\int_{\Gamma_1} \frac{\partial E_\alpha}{\partial \nu} v \ \ ds -
\int_{\partial \Omega} \frac{\partial v}{\partial \nu} E_\alpha \
\ ds &=&\ds  \alpha^d \sum_{j=1}^m (1 - \frac{\mu_j}{\mu_0})
\nabla E_0(z_j) \cdot M_j(\frac{\mu_j}{\mu_0})  \nabla v(z_j) \\
 \nm
 &+& \ds  \alpha^d k^2 \sum_{j=1}^m (1 - \frac{\eps_j}{\eps_0})
|B_j| E_0(z_j) v(z_j) + o(\alpha^d),
\end{array}
\ee where $|B_j|$ stands for the volume of the set $B_j$. We want
to make suitable
choices for the test functions
$v$ in $\tilde{N}(\Omega)$ and the boundary condition $E_\alpha
|_{\partial \Omega}$ in order to get simple
equations for the unknown parameters, namely, for
the points
$\{z_j\}_{j=1}^m$ and matrices $\{M_j\}_{j=1}^m$. Similar
idea was used and the associated numerical experiments have been
successfully
conducted in the case of the conductivity problem \cite{AMV} with
boundary measurements on all of $\partial \Omega$.

Let us describe
our inversion method. Take $\eta$ to be a vector in $\R^d$,
$\eta^{\perp}$ a unit vector in $\R^d$ which is orthogonal to
$\eta$, and $\gamma$ a complex number. Then $e^{i(\eta+ \gamma
\eta^{\perp})\cdot x}$ is a solution to the Helmholtz equation in
$\R^d$ if and only if $\gamma^2 = k^2 - | \eta |^2$ and in this
case $e^{i(\eta - \gamma \eta^{\perp})\cdot x}$ is also a solution
to the Helmholtz equation in $\R^d$. For simplicity, let us
consider the case where all the $B_j$ are balls. In this case all
the matrices $M_j$ are  multiples of the identity matrix which
makes our analysis simpler.

If $\frac{\partial E_\alpha}{\partial \nu}$ is known on the whole
boundary $\partial \Omega$ then taking $E_\alpha = e^{i(\eta+
\gamma \eta^{\perp})\cdot x}$ on $\partial \Omega$ we know from
\cite{AMV} that
\[\begin{array}{l} \ds
\int_{\partial \Omega} \frac{\partial E_\alpha}{\partial \nu}(y)
e^{i(\eta -  \gamma \eta^{\perp})\cdot y}\ \ ds(y) -
\int_{\partial \Omega} \frac{\partial}{\partial \nu}(e^{i(\eta -
\gamma \eta^{\perp})\cdot y}) E_\alpha(y) \ \ ds(y)
\\ \nm \ \ \ \ = \ds \alpha^d \sum_{j=1}^m e^{2 i \eta \cdot z_j}
\Bigr[ (1 - \frac{\mu^j}{\mu^0}) M_j(\frac{\mu^j}{\mu^0}) (2 |
\eta|^2 - k^2) + \ds  k^2 (1 - \frac{\eps^j}{\eps^0}) |B_j| \Bigr]
+ o(\alpha^d).
\end{array}
\]
The main difficulty in generalizing this approach to the case when
$\frac{\partial E_\alpha}{\partial \nu}$ is known only on a part
$\Gamma_1 \subset \subset \partial \Omega$ is to construct a
function $w_\alpha(x)$ in $\tilde{N}(\Omega)$,  that is
asymptotically $e^{i(\eta -  \gamma \eta^{\perp})\cdot x}$ as
$\alpha$ approaches $0$. The following lemma holds.
\begin{lem} \label{lem1} Let $\Omega^\prime \subset \subset \Omega$ be
a $C^2$-domain. Let $\eta \in \R^d$ and $\eta^{\perp}$ be a
unit vector in $\R^d$ that is orthogonal to $\eta$. There exists
$w_\alpha \in \tilde{N}(\Omega)$ such that
\[ w_\alpha(x) = e^{i(\eta -  \gamma
\eta^{\perp})\cdot x} + o(\alpha^d) \ \mbox{and} \ \nabla
w_\alpha(x) = i(\eta -  \gamma \eta^{\perp}) e^{i(\eta -  \gamma
\eta^{\perp})\cdot x} + o(\alpha^d),
\]
uniformly in $\Omega^\prime$.
\end{lem}
This lemma is an immediate corollary of the following general
density result.
\begin{prop} \label{lem2}
The set $\tilde{N}(\Omega)$ is dense, in the $L^2(\Omega^\prime)$
norm, in the set  $N(\Omega)$.
\end{prop}
\proof Assume the contrary and let $v \in N(\Omega)$ be an element which
cannot be
approximated in $L^2(\Omega^\prime)$ by the functions from
$\tilde{N}(\Omega)$ with a prescribed accuracy. Then
there is an element in $N(\Omega)$, which we denote again $v$, such that
$\int_{\Omega^\prime}
v w dx = 0, \forall
\; w \in  \tilde{N}(\Omega)$. Let $G_0$ be the Dirichlet Green's
function in $\Omega$: $$\left\{
\begin{array}{l}
\Delta G_0 + k^2 G_0 = \delta_y(x) \ \ \mbox{in} \ \ \Omega,\\ G_0
= 0 \ \ \mbox{on} \ \ \partial \Omega.
\end{array}
\right.
 $$
Define $\tilde{H}^{3/2}(\Gamma_1) = \{ p \in H^{3/2}(\Gamma_1) \
\mbox{such that there exists} \ \tilde{p}  \in H^{3/2}(\partial \Omega),
\tilde{p}|_{\Gamma_2}  = 0,\tilde{p}|_{\Gamma_1}  = p \}$.  
 Since any $w \in \tilde{N}(\Omega)$ can be represented as
follows $$w(x) = \int_{\Gamma_1} \frac{\partial G_0(x,y)}{\partial
\nu(y)} \ p(y) \ ds(y), x \in \Omega,$$ where $p \in
\tilde{H}^{3/2}(\Gamma_1)$ is arbitrary, we have $$\int_{\Omega^\prime} v(y)
\frac{\partial G_0(x,y)}{\partial \nu(x)} \ dy =0, \ \forall \ x
\in \Gamma_1.$$ Define $u(x): = \int_{\Omega^\prime} v(y)G_0(x, y)
\ dy.$ Then $u|_{\partial \Omega} = 0, \frac{\partial u}{\partial
\nu} |_{\Gamma_1} = 0$, and $$\Delta u + k^2 u = \left\{
\begin{array}{l} v \ \  \mbox{in } \Omega^\prime,\\ 0  \ \
\mbox{in } \Omega \setminus \overline{\Omega^\prime}.
\end{array}
\right. $$ By the uniqueness of the solution to the Cauchy problem
for the Helmholtz equation, it follows that $u = 0$ in $\Omega \setminus
\overline{\Omega^\prime},$ and so $u =  \frac{\partial u}{\partial
\nu}  = 0$ on $\partial \Omega^\prime$. Since $\Delta u + k^2 u =
v $ in $\Omega^\prime$, it follows by multiplying this equation by
$v$ and  integrating by parts over $\Omega^\prime$ that $
\int_{\Omega^\prime} v^2 dx= 0.$ Thus, $v=0$ in $\Omega^\prime$,
and, by the unique continuation, $v=0$ in  $\Omega,$ which
proves Proposition 2.1. In the above argument we assumed that $u$
and $v$ are real valued. This can be done without loss of
generality since $k^2>0$.
 \square

From Proposition \ref{lem2} it follows that the function
$w_\alpha$ in Lemma \ref{lem1} can be constructed (numerically) by
solving the minimization problem: \be \label{f4} \min_{p \in
\tilde{H}^{3/2}(\Gamma_1)} \| \int_{\Gamma_1} \frac{\partial
G_0(x,y)}{\partial \nu(y)} \ p(y) \ ds(y) - e^{i(\eta -  \gamma
\eta^{\perp})\cdot x} \|_{L^2(\Omega^\prime)}. \ee
We can take $$
w_\alpha(x) = \int_{\Gamma_1} \frac{\partial G_0(x,y)}{\partial
\nu(y)} \ p_\alpha(y) \ ds(y),$$ where $p_\alpha$ is any function
in $\tilde{H}^{1/2}(\Gamma_1)$ such that \be \label{eq4}\| \int_{\Gamma_1}
\frac{\partial G_0(x, y)}{\partial \nu(y)} \ p_\alpha(y) \  ds(y)
- e^{i(\eta - \gamma \eta^{\perp})\cdot x} \|_{L^2(\Omega^\prime)}
=o(\alpha^d). \ee
 Since $(\Delta + k^2) (w_\alpha -e^{i(\eta -
\gamma \eta^{\perp})\cdot x}) = 0$ in $\Omega$, by the standard
elliptic interior estimates \cite{GT}, we obtain from (\ref{eq4})
that $\|w_\alpha -e^{i(\eta - \gamma \eta^{\perp})\cdot
x}\|_{H^s(\Omega^{\prime \prime})}$ is of order $o(\alpha^d)$ for any
$s$ and any $C^2$-domain $\Omega^{\prime \prime} \subset \subset
\Omega^\prime$. Lemma \ref{lem1} follows then immediately from the
Sobolev imbedding theorems.

 Now, if we choose $E_\alpha = e^{i(\eta+ \gamma \eta^{\perp})\cdot
x}$ on $\partial \Omega$ and $v = w_\alpha$ in $\Omega$ then,
since the points $\{z_j\}_{j=1}^m$ are away
from the boundary $\partial \Omega$, it follows from (\ref{eq3})
and Lemma \ref{lem1} that the following asymptotic expansion holds:
\be
\label{f3}
\begin{array}{lll} \ds \Lambda_\alpha(\eta) &=&\ds
\int_{\Gamma_1} \frac{\partial E_\alpha}{\partial \nu} w_\alpha  \
ds - \int_{\partial \Omega} \frac{\partial w_\alpha}{\partial \nu}
E_\alpha  \ ds \\ \nm &=&\ds \alpha^d \sum_{j=1}^m e^{2 i \eta
\cdot z_j} \Bigr[ (1 - \frac{\mu_j}{\mu_0})
M_j(\frac{\mu_j}{\mu_0}) (2 | \eta|^2 - k^2) + \ds  k^2 (1 -
\frac{\eps_j}{\eps_0}) |B_j| \Bigr] + o(\alpha^d).
\end{array}
\ee
 Recall that $e^{2 i \eta \cdot z_j}$ (up to a
multiplicative constant) is  the Fourier transform of the
Dirac function $\delta_{-2 z_j}$ (a point mass located at $-2
z_j$). Multiplication by powers of $\eta$ in the Fourier space
corresponds to differentiation of the Dirac function.  Therefore, the
function  $\Lambda_\alpha(\eta)$ is the inverse Fourier
transform of a distribution supported at the points
$z_j$. A numerical
Fourier inversion of a sample of $\Lambda_\alpha(\eta)$ will
yield the points $z_j$ with a small error as
$\alpha \to 0$. It is natural to use a
fast
Fourier transform for this inversion.
One can estimate the number of the
sampling points needed for an accurate discrete Fourier
inversion  using the Shannon's theorem \cite{D}, page 18.
This number is
 of order $\ds (\frac{h}{\delta})^3$. One needs this amount of sampled
values of  $\eta$ to reconstruct, with resolution $\delta$, a
collection of inhomogeneities that lie inside a square of side
$h$. Once the points $\{z_j\}_{j=1}^m$ are found, one may find
$\{M_j\}_{j=1}^m$ by solving appropriate linear system arising
from the asymptotic formula (\ref{eq3}). If $B_j$ are general
domains, our calculations become more complicated, and eventually we
have to deal with pseudo-differential operators (independent of
the space variable $x$) applied to the same Dirac functions.
 In view of the asymptotic results derived in \cite{AVV} a
 similar approach may be applied to the full Maxwell equations in the presence of
 small dielectric inhomogeneities.

\section*{Acknowledgements}
This work is partially supported by ACI Jeunes Chercheurs (0693)
from the Ministry of Education and Scientific Research, France.

\end{document}